\documentstyle[aps,prl,twocolumn,epsf,rotate]{revtex}

\newcommand{\kk}{{\bf k}}
\newcommand{\RR}{{\bf R}}
\newcommand{\qq}{{\bf q}}
\newcommand{\QQ}{{\bf Q}}

\newcommand{\dd}{{\vec{\bf \delta}}}
\newcommand{\cc}{{\hat c}}
\newcommand{\ah}{{\hat a}}

\newcommand{\ccd}{{\hat c^\dagger}}

\begin{document}
\twocolumn[\hsize\textwidth\columnwidth\hsize\csname
@twocolumnfalse\endcsname

\title{2D bands and electron-phonon interactions in polyacene plastic transistors}

\draft
\author{Jairo Sinova, John Schliemann, 
Alvaro S. N\'{u}\~{n}ez, and A. H. MacDonald}
\address{Department of Physics,
University of Texas, Austin, Texas 78712-1081}
\date{\today}
\maketitle

\begin{abstract}

We present a simple tight-binding model for the two-dimensional energy bands of 
polyacene field-effect transistors and for the coupling of these
bands to lattice vibrations of their host molecular crystal. 
We argue that the strongest electron-phonon interactions in these systems originate from the   
dependence of inter-molecule hopping amplitudes on collective molecular motion,
and introduce a generalized Su-Schrieffer-Heeger model that accounts for all vibrations  
and is parameter-free once the band mass has been specified.
We compute the electron-phonon spectral function $\alpha^2F(\omega)$ as a 
function of two-dimensional  hole density, and are able to explain the onset of  
superconductivity near 2D carrier density $n_{2D} \sim  10^{14} {\rm cm}^{-2}$, 
discovered in recent experiments by Sch\"on {\it et al.} \onlinecite{Batlogg}.

\end{abstract}


\vskip2pc]

Recent studies\cite{Batlogg,BatloggPRL}
of high-mobility organic field-effect transistors by Batlogg and collaborators 
have demonstrated that their electronic properties can be tuned 
through an astonishing range, simply by adjusting the two-dimensional (2D)
electron density $n_{2D}$ with a gate voltage.  The high quality of these organic single crystals   
and of their interfaces with AlO$_{3}$ dielectrics opens up new possibilities for studying
the physics of coherent {\em band} quasiparticle transport in organic semiconductors.
To date, however, 
the analysis of organic field-effect-transistor electronic systems has been hampered by the absence of 
simple and reliable models for their electronic quasiparticles and for the interactions of 
these quasiparticles with each other and with vibrations of the host lattice.  Indeed the discovery of 
coherent band quasiparticle properties and of the quantum Hall effect in these
systems, which are usually thought of as being complex and relatively disordered, 
has been one of the major surprises that has emerged from recent materials quality advances.

The electronic properties of polyacene semiconductors are normally described by 
{\em small polaron} theory\cite{Kenkre}, which is quite successful at temperatures
above $\approx $20 K.   Small polaron theory starts from localized molecular orbitals  
and is therefore unable to describe the low-temperature
{\em band} quasiparticle behavior seen in Battlogg {\em et al.'s} samples, which have 
room-temperature to low temperature resistance ratios $\sim 10^{4}$.
Schubnikov-de Haas transport studies in these samples 
establish quasiparticle mean-free-paths in excess of 1000 lattice constants,
demonstrating that 
the low-temperature regime can be described only by starting at the opposite limit and building 
a theory based on delocalized band electron states.  Nevertheless, the interactions of quasiparticles
with lattice vibrations remain strong, and are presumably responsible for superconductivity.
In this Letter, we propose a simple 
tight-binding model for the quasiparticle bands that emerge from $\pi$-molecular orbitals 
and are filled with carriers by the field-effect, and  
for the interaction of these bands with the host crystal vibrational excitations.
Out theory is parameter free once the band mass has been specified \cite{BatloggPRL}.

We find that over the entire wide range of studied densities, 
the quasiparticles of these systems lie within a single 2D 
band, bound to the crystal surface by the transistor's electric field.  
We argue that the dominant 
electron-phonon interactions in polyacene crystals differ qualitatively from
the deformation and Frohlich interactions of inorganic semiconductors\cite{DasSarma}
and from the intra-molecule interactions that dominate in doped fullerenes\cite{fullerences}, 
being more similar instead to those of doped polyacetelene\cite{SSH} crystals.
We describe these interactions using a
Su-Schrieffer-Heeger model\cite{SSH}, detailed in the following paragraphs, 
generalized to account for the larger number of important rigid molecule motions. 
The lattice phonons, a mixture in general of translational and librational components, 
induce a change in the value of the hopping integral and hence couple\cite{SSH} to the quasiparticle
bands.  We estimate the variation of hopping parameters with vibrational normal coordinates
by assumming that they are proportional to the spatial overlap of  
H\"{u}ckel-type LUMO and HOMO molecular orbitals, with the proportionality constant fixed by 
matching the measured equilibrium hopping parameter \cite{Lannoo2}.  
In this way, we have evaluated the carrier density dependence of $\alpha^2F(\omega)$, the 
index of electron-phonon interactions that figures 
prominently in the theory of phonon-mediated superconductivity.
We find peaks that match the spectrum extracted from recent tunneling experiments\cite{BatloggPRL}.  
We also find that $\alpha^2F(\omega)$ is large, due to a density-of-states
peak, when the lowest energy 2D subband is close to half-filling.  
We propose this property as the explanation of the relatively sharp onset
of superconductivity at a 2D carrier density of $\approx 10^{14}{\rm cm^{-2}}$
seen in experiments\cite{Batlogg}.

We focus in this study on anthracene which is monoclinic ($P2_{1/a}$)
with two basis molecules whose orientations are related by a 
gliding plane symmetry \cite{anth_parameters}.  
Since Huckel model intra-molecular hopping parameters have no linear dependence 
on low energy intra-molecular vibration normal coordinates, our model Hamiltonian 
includes only inter-molecular hopping parameters \cite{lannoo}.
\begin{eqnarray*}                                                                 
{\cal H}&=&{\cal H}_{\rm t}+{\cal H}_{\rm e-e}   
+{\cal H}_{\rm vib}+{\cal H}_{\rm ext}           
\end{eqnarray*}                                                                   
where ${\cal H}_{\rm t}$ is the LUMO or HOMO band Hamiltonian described by a 
nearest-neighbour hopping model, 
${\cal H}_{\rm e-e}$ describes Coulomb interaction between the electons, 
${\cal H}_{\rm vib}$ is the free-phonon Hamiltonian, 
and ${\cal H}_{\rm ext}$ describes the external electric field from the transistor's gate.

\begin{figure}
\epsfysize=2.70in
\epsfxsize=3.00in
\centerline{\epsffile{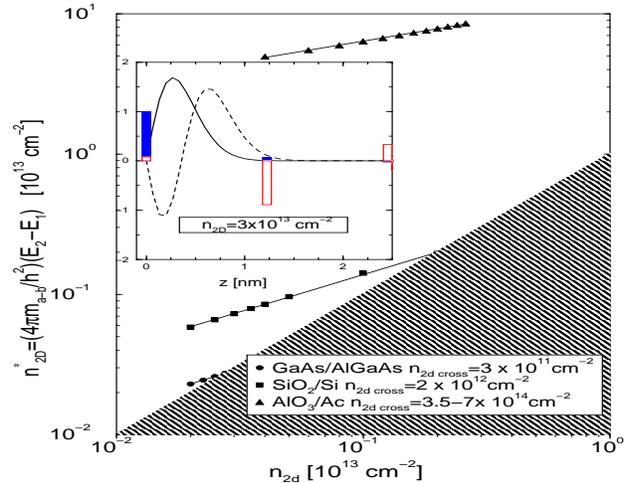}}
\caption{States per area with energy below the bottom of the second subband {\it vs.}
2D electron density for GaAs, Si, and polyacene field effect transistors from envelope-function
LDA calculations.  Second subband occupation occurs in the hatched area. 
The inset shows  first and second subband 
wavefunctions obtained from envelope function LDA 
and tight-binding model Hartre-approximation calculations (bars) at $n_{2D}= 3
\times 10^{13}{\rm cm}^{-2}$.}
\label{LDA}
\end{figure}

We start by considering the field-induced energy bands obtained neglecting electron-phonon
interactions.  The key question we need to address is 
the number of 2D subbands that are occupied at a 
particular 2D density, $n_{2D}$. 
As summarized in Fig.~\ref{LDA}, we find that all carriers reside in a single
subband up to much higher density in these transistors than in their inorganic counterparts,
principally because of the relatively small dielectric constants 
($\epsilon\sim 3.5$) 
and the large effective in plane band mass ($m_\perp=1.5 m_e$) and 
in spite of the rather large c-direction effective band mass $m_z\sim 3 m_e$ \cite{Batlogg,note2}. 
Since the Al$_2$O$_3$ barrier is quite high
($\approx 1.3\, {\rm eV}$), we take it to be infinite.
Our conclusion is based partially on envelope-function density-functional calculations
for which the local-density-approximation (LDA) can be used to estimate many-body 
effects\cite{note3,Ando_Fowler_Stern} that favor a single subband.  
In our LDA calculations we find that the second subband
is first populated at $n_{2D} \approx 5-7\times 10^{14}{\rm cm}^{-2}$, a density just 
larger than the highest achieved in experimental systems.  However, as shown in the 
inset of Fig.~\ref{LDA}, the wavefunctions in these calculations are 
already localized over $\sim 1-2$ nm at densities well below this maximum value.
We have therefore repeated these electronic structure calculations using our
tight-binding model, and treating interactions in a Hartree approximation.  With this approach
we find that $99\%$ of the charge density is in the first molecular layer at $n_{2D}= 3
\times 10^{13}{\rm cm}^{-2}$, the density at which second subband occupation first occurs
in this approximation. 
Since we believe that the experimental signatures of second subband
occupation would be unambiguous and no anomalies in 2D density dependence have been reported,
we conclude that a single 2D subband is occupied up to the highest densities and that the 
lowest subband is strongly localized in the top layer for $n_{2D}$ larger than $\approx 10^{13} 
{\rm cm}^{-2}$. 

To model the electron-phonon interaction, we expand each $a-b$ plane hopping 
parameter to first order in the twelve coordinates that describe rigid rotations and 
displacements of neighboring polyacene molecules:
\begin{eqnarray}
t&=&t_0+ \sum_{m=1}^6\sum_{\mu}\tilde{\bf t}_{\mu,m}
\tilde{u}_{\mu,m}
\label{hop_integral}
\end{eqnarray}
where $\tilde{u}_{\mu,m}$ is the generalized displacement coordinate,
$\mu =1,2$ is the molecular basis index, $m=1,2,3$ denote 
displacements of the molecular center of mass along
$\hat x$,$\hat y$ and $\hat z$ directions, and $m=4,5,6$ denote
angular displacements around the $1,2,3$ principal axes 
of each molecule. The electron-phonon interaction parameters 
$\tilde{\bf t}_{\mu,m}\equiv \partial t /\partial \tilde{u}_{\mu,m}$.  
are calculated based on the assumption of 
proportionality between hopping integrals and overlap integrals between HOMO or LUMO 
orbitals on adjacent molecules. 
The values obtained using standard 
H\"{u}ckel approximation\cite{Lannoo2} HOMO $\pi$ orbitals 
(and hence appropriate to the hole systems on which we focus)
are listed in Table I in units of $t_0/a$ for $m=1,2,3$ and $t_0$ for
$m=4,5,6$. 
\vspace{-0.4cm}

\begin{figure}
\epsfxsize=3.40in
\epsfysize=3.10in
\centerline{\epsffile{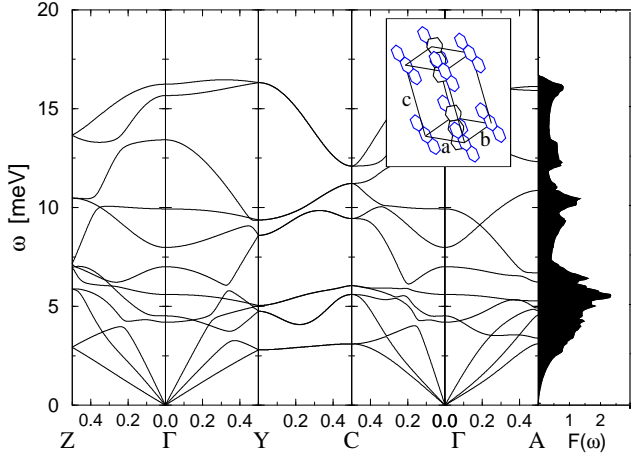}}
\caption{Calculated dispersion curves in anthracene (structure shown in inset). Shown on the
right is the phonon density of states $F(\omega)$ in units of
${\rm meV}^{-1}$ and normalized to 12, the number of vibrational
modes considered.}
\label{dispersion}
\end{figure}

We now derive expressions for the interactions between 
phonons of the host crystal and the 2D band quasiparticles
whose Hamiltonian is:  
\begin{eqnarray}
{\cal H}^0_{2D}&=&
\sum_{\alpha \kk\in{\rm BZ_{2D}}}
\sum_{i=1,2} (\epsilon_{i,\alpha}(\kk)+E_\alpha-\mu)
\ccd_{i,\kk}\cc_{i,\kk}
\end{eqnarray}
with
\begin{eqnarray}
\epsilon_{i}(\kk)&=&
(-1)^i 2t_0\left[\cos(\frac{k_x a+k_y b}{2})+\cos(\frac{k_x a-k_y b}{2})
\right] 
\end{eqnarray}
where $E_\alpha$ is obtained from the discrete layer subband calculation,
the two electronic band indices originate from the two molecules per unit cell,
and ${\rm BZ_{2D}}$ indicates the two dimensional rectangular lattice Brillouin zone.
The generalized displacements $\tilde{u}_{\mu,m}$ in Eq. \ref{hop_integral} 
can be written in terms of quantized vibrational normal modes as follows: 
\begin{eqnarray*}
&\tilde{u}_{\mu,m}(\RR)=\sum_{\QQ\in{\rm BZ},\nu}\sqrt{\frac{\hbar}{2
\tilde{M}_{\mu,m} N\omega_\nu(\QQ)}}\times\\&
\left[ {\epsilon}_{\mu,m}(\QQ,\nu) \ah_{\QQ,\nu} e^{i \QQ\cdot\RR}+
\epsilon_{\mu,m}^*(\QQ,\nu) \ah^\dagger_{-\QQ,\nu}
e^{-i\QQ\cdot\RR}\right]
\end{eqnarray*}
Here $N$ is the total number of  unit cells,
$\tilde{M}_{\mu m}$ is the molecular mass
of the $\mu$th molecule for $m=1,2,3$ and
the molecular moment of inertia around the
$(m-3)'$th principal axis $I^\mu_{(m-3)}$ for m=4,5,6.
The phonon frequencies and polarization vectors are obtained
by solving the standard secular equation
\begin{eqnarray}
\omega_i^2(\qq) \epsilon_m^i(\qq,\mu)=
\sum_{\nu,n} D_{\mu m,\nu n}(\qq) \epsilon^i_n(\qq,\nu)
\end{eqnarray}
These eigenmodes are in general 
a mixture of displacements and rotations\cite{note1}. 
We compute the dynamical matrix following the procedure outlined
in Ref. \cite{Taddei_Dorner}. 
The phonon density-of-states and dispersion curves that emerge from these 
calculations are shown in Fig. \ref{dispersion}.
The rigid-molecule vibration restriction used here is a convenient but inessential approximation that 
can be complemented if necessary by including coupling to important isolated molecule vibrations; 
it is however
reasonably accurate for low-frequency vibrations of anthracene becoming less reliable for larger 
polyacenes.
\vspace{-1.00cm}

Combining these ingredients we finally rewrite the Hamiltonian 
(treating Coulomb interactions at the mean field level) as 
\begin{eqnarray}
{\cal H}={\cal H}_{\rm 2D}^0+{\cal H}_{\rm 2De-vib}+{\cal H}_{\rm vib}
\end{eqnarray}
where
\begin{eqnarray}
{\cal H}_{\rm 2De-vib}&=&
\frac{1}{\sqrt{N}}\sum_{\kk\in{\rm BZ}}\sum_{\nu\alpha\beta,i,j}
\sum_{\QQ\in {\rm BZ}}
g_{i,\alpha;j,\beta}(\kk,\QQ,\nu)\times\nonumber\\&&
[\ah_{\QQ,\nu} +\ah^\dagger_{-\QQ,\nu}]\ccd_{i,\alpha [\kk+\qq]}
\cc_{j,\beta \kk}
\label{e-ph}
\end{eqnarray}
and
\begin{eqnarray}
&&g_{i,\alpha;j,\beta}(\kk,\qq,q_z,\nu)=
(-1)^i\frac{\tilde{f}_{\alpha,\beta}(q_z)}{2}
\sum_{\dd',\mu,m} \sqrt{\frac{\hbar}{2
\tilde{M}_{\mu,m} \omega_\nu(\QQ)}}\nonumber\\&&
\times\tilde{\bf t}_{\mu m}(\dd')\epsilon_{\mu m}(\QQ,\nu)
e^{i \qq\cdot\dd_\mu}(e^{-i\dd'\cdot[\kk+\qq]}
+(-1)^{(i-j)}e^{i\dd'\cdot\kk}).
\label{Tij}
\end{eqnarray}
In Eq.~\ref{Tij} $\tilde{f}_{\alpha,\beta}(q_z)=\sum_{z_l}a^{\alpha *}_l a^{\beta}_{l}
e^{iq_z z_l}$ is the form factor of the subband tight-binding wavefunction
and the brackets indicate reduction to the 2D BZ. In our 
calculations we use the strict 2D limit 
($\tilde{f}_{\alpha,\alpha}(q_z)=1$) because of the strong quantum
confinement found in  LDA and Hartree calculations.

\begin{figure}
\epsfxsize=3.in
\epsfysize=2.75in
\centerline{\epsffile{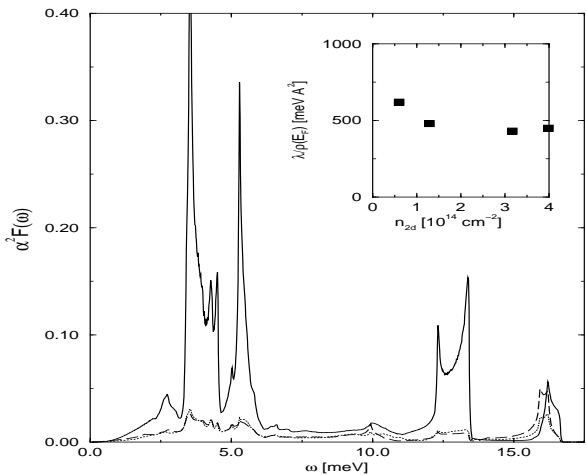}}
\caption{$\alpha^2F(\omega)$ for $n_{\rm 2d}=4.0 \times 10^{14}
\,{\rm cm}^{-2}$ (solid line), 
$n_{\rm 2d}=1.4 \times 10^{14} \,{\rm cm}^{-2}$ (dotted line),
and $n_{\rm 2d}=6.7 \times 10^{13} \,{\rm cm}^{-2}$ (dashed line).}
\label{alpha2F}
\end{figure}

We define the electron-phonon interaction spectral function in the standard way:
\begin{eqnarray}
&\alpha&^2F(\omega)\equiv \frac{1}{N g(\mu)}
\sum_{\kk,\kk',i}\sum_{q_z,\nu} |g_{i,1;i,1}(\kk,[\kk'-\kk],q_z,\nu)|^2 
\nonumber\\&\times&\delta(\epsilon_i(\kk)-\mu) \delta(\epsilon_i(\kk')-\mu)
\delta(\omega-\omega_\nu([\kk'-\kk],q_z)
\end{eqnarray}
where $g(\mu)=\sum_\kk \delta(\epsilon_i(\kk)-\mu)$ is the electronic density of 
states, and $[\kk'-\kk]$ denotes the projection of the $\QQ=(\kk'-\kk,q_z)$
vector into the three dimensional BZ.   The results of these calculations 
for densities $n_{\rm 2d}=4.0 \times 10^{14}
\,{\rm cm}^{-2}$,
$n_{\rm 2D}=1.4 \times 10^{14} \,{\rm cm}^{-2}$,
and $n_{\rm 2D}=6.7 \times 10^{13} \,{\rm cm}^{-2}$ are 
shown in Fig. \ref{alpha2F}. The peak locations in 
the low-frequency rigid molecule vibration regime agree with those
observed in infrared absorption and tunneling data in pentacene
crystals \cite{BatloggPRL}. 
The mass enhancement factor $\lambda$, obtained from the integration of
$2F\alpha^2(\omega)/\omega$, close to half filling is $0.25$. Its sharp decay
away from half filling (driven by the density of states) helps explain the sharp 
superconductivity onset observed in the experiments. 
We estimate the supercunducting critical temperature by the
[dimension independent] BCS expression $k_B T_c\approx \hbar\omega_{D} \exp[-1/\lambda]$. 
Such expression, using $\hbar\omega_{D}/k_B\approx 150 K$ from Ref. \cite{BatloggPRL}, yields $\sim 3 K$
in agreement with the experiments\cite{Batlogg}. 
We note that the inclusion of the second-nearest-neighbour hopping tends to push the 
density of states peak to higher energies (densities). Adding a finite life time of the 
quasiparticles, will broaden these peaks, hence creating a plateau
in the  n$_{\rm 2D}$-dependence of $T_c$ shifted to
the right of half filling as observed in the experiments \cite{Batlogg}.
This last effect would lower also the  $T_c$ calculated at $n=4.00 \times 10^{-12}{\rm cm}^{-2}$,
worsening experimental agreement.
However, given that $E_F$ does not satisfy $E_F>>\hbar\omega_{max}$, such estimates of $T_c$ from
$\lambda$ are qualitative.

In summary we have presented a theory of the low-temperature quasiparticle
bands in polyacene field effect transistors, and of the coupling of these bands to
the host molecular crystal phonons.  We predict that the quasiparticles lie in a 
single 2D tight-binding band up to the highest densities that have been achieved 
experimentally at present and that the most important electron-phonon interactions 
arise from the influence of approximately rigid molecular translations and 
rotations on hopping between HOMO and LUMO orbitals on adjacent molecules.  
This picture implies a dependence of superconducting critical temperature on
molecular lattice constant which contrasts with the case of doped fullerene 
superconductors.  For the fullerenes, the important electron-phonon interactions 
are intramolecular so that $T_c$ depends on intermolecular hopping only 
through the density-of-states.  Decreasing the lattice constant, increases
hopping, and hence decreases the density-of-states and $T_c$\cite{fullerences}.  Here decreasing 
the lattice constant will also strengthen the important electron-phonon interactions.
According to our theory, the latter effect dominates and $T_c$ will increase with 
decreasing lattice constant.  We find that phonon-mediated electron-electron
interactions in polyacene molecular crystals are strong only when the 2D tight-binding
band is close to half-filling and its density of states is relatively large.
This occurs for 2D densities comparable to $4 \times 10^{14} {\rm cm}^{-2}$, where
superconductivity turns on relatively abruptly.  At lower carrier densities,
$\lambda \ll 1$ and electron-electron interactions are dominated by repulsive
Coulomb interactions, consistent with the occurrence of the fractional quantum Hall effect.

The authors acknowledge helpful discussions with P. Barbara, 
B. Batlogg, A. Dodabalapur,  Y. Joglekar, T. Jungwirth, and P. Rossky.
This work was supported by the Deutsche
Forschungsgemeischaft, by the Welch Foundation and the by the National Science
Foundation under grant DMR0115947.

\begin{table*}
\begin{tabular}{|c|c|c|}
$\delta_1=(a/2,b/2)$& $\delta_1=(a/2,b/2)$&$ \delta_2=(-a/2,-b/2)$\\\hline\hline
$\tilde{\bf t}_{11}=2.2706$&$\tilde{\bf t}_{14}=1.4895$&$\tilde{\bf t}_{14}=-1.6095$\\\hline
$\tilde{\bf t}_{12}=3.3842$&$\tilde{\bf t}_{15}=2.0318$&$\tilde{\bf t}_{15}=1.8288$\\\hline
$\tilde{\bf t}_{13}=7.5308$&$\tilde{\bf t}_{16}=1.0082$&$\tilde{\bf t}_{16}=1.4658$
\end{tabular}
\caption{Electron-phonon interaction parameters.  Note that interaction parameters for 
different near neighbors are different.  All interaction parameters are related by 
symmetry to the nine values list above.  A complete list of hopping parameters for all
near neighbors of the two molecules in 
the crystal's primitive cell is available from the authors.}
\end{table*}


\begin{references}

\bibitem{Batlogg} 
J. H. Sch$\rm\ddot{o}$n, Ch. Kloc, and B. Batlogg, Science {\bf 288}, 2338 (2000);
J. H. Sch$\rm\ddot{o}$n {\it et al}.,
Science {\bf 287}, 1022 (2000); 
J. H. Sch$\rm\ddot{o}$n {\it et al}., 
Science {\bf 289}, 599 (2000);
J. H. Sch$\rm\ddot{o}$n, Ch. Kloc, and B. Batlogg,
Nature {\bf 406}, 702 (2000).

\bibitem{BatloggPRL} Mark Lee {\it et al}., 
Phys. Rev. Lett. {\bf 86}, 862 (2001).

\bibitem{Kenkre} V.M. Kenkre {\it et al}., 
Phys. Rev. Lett. {\bf 62}, 1165 (1987). 

\bibitem{DasSarma} S. Das Sarma and B. A. Mason, Annals of Physics
{\bf 163}, 78 (1985).

\bibitem{fullerences} M. Schl\"uter {\it et al}., 
Phys. Rev. Lett. {\bf 68}, 526 (1992);
C.M. Varma, J. Zaanen, and K. Raghavachari, Science {\bf 254}, 989 (1991);
For a review see O. Gunnarsson, Rev. Mod. Phys. {\bf 69}, 575 (1997).

\bibitem{SSH} W. P. Su, J. R. Schrieffer, and A. J. Heeger,
Phys. Rev. Lett. {\bf 42}, 1698 (1979); W. P. Su, J. R. Schrieffer,
and A. J. Heeger, Phys. Rev. B {\bf 22}, 2099 (1980).

\bibitem{Lannoo2} Orbitals obtained through the Huckle approximation are in
good agreement with the ones obtained through more sophisticated LDA calculations. See
for example Ref. \cite{fullerences}. 

\bibitem{anth_parameters} Here the crystal parameters are
$a=8.378$ \r{A}, $b=5.981$ \r{A}, $c=11.059$ 
\r{A}, and $\beta=125.34$\r{} are the parameters
at 4.7 K obtained from Dorner {\it et al} [13].

\bibitem{lannoo} Higher frequency breathing modes of the isolated 
molecule will result in shifts of LUMO and HOMO levels, and to electron-phonon
interactions that are not included in our model.  However, these coupling are 
expected to be substantially screened in a molecular crystal and there is no
%
evidence for their presence in tunneling studies of the polyacene superconducting state.
For a discussion of this contribution see A. Devos and M. Lannoo,
Phys. Rev. B {\bf 58}, 8236 (1998). 

\bibitem{note2} In the triangular well approximation the $n_{2D}$ at which the second subband
occupation begins is proportional to $(m_{\perp}/m_z)^3 (m_z/\epsilon)$.

\bibitem{note3} Here we use anisotropic 3D electron gas with $m_z=m_{\perp}=1.5 m_e$ to generate
the LDA correlation potential.

\bibitem{Ando_Fowler_Stern} T. Ando,A. B. Fowler,and F. Stern, Rev. Mod. Phys.  54, 437 (1982). 


\bibitem{note1}Intra-molecular motion can be added to the model simply by adding 
the appropriate twist and butterfly vibrations to the list of 
generalized coordinates, inserting the isolated molecule vibration frequencies
as diagonal contributions to the augmented dynamical matrix, 
and including the influence of these displacements on
inter-molecule interactions.  Note that when this is done, a general vibrational mode
couples rigid molecule and intra-molecule generalized displacements.

\bibitem{Taddei_Dorner} G. Taddei {\it et al}., 
J. Chem. Phys. {\bf 58}, 966 (1973); 
B. Dorner {\it et al}., 
J. Phys. C {\bf 15}, 2353 (1982).



\end{references}
\end{document}